# An EMG study of the lip muscles during covert auditory verbal hallucinations in schizophrenia

Lucile Rapin[1,2], Marion Dohen[2], Mircea Polosan[3], Pascal Perrier[2], Hélène Lœvenbruck[2]

[1]Département Education et Pédagogie, Université du Québec à Montréal, Québec
[2]GIPSA-lab, Speech and Cognition Department, UMR 5216, CNRS, Université de Grenoble, France
[3]Pôle de Psychiatrie et de Neurologie, CHU de Grenoble, France

Corresponding author:
Hélène Lœvenbruck
GIPSA-lab, Département Parole et Cognition
and
Laboratoire de Psychologie et NeuroCognition
CNRS UMR 5105 Université de Grenoble-Alpes
Bâtiment Sciences de l'Homme et Mathématiques BP47
38040 Grenoble Cedex 9 France
Helene.Loevenbruck@upmf.-grenoble.fr
Tel: +33 4 76 82 56 75
Fax: +33 4 76 82 78 11

Word counts:
Abstract: 198

*Abstract*

*Purpose:* Auditory verbal hallucinations (AVHs) are speech perceptions in the absence of a external stimulation. An influential theoretical account of AVHs in schizophrenia claims that a deficit in inner speech monitoring would cause the verbal thoughts of the patient to be perceived as external voices. The account is based on a predictive control model, in which verbal self-monitoring is implemented. The aim of this study was to examine lip muscle activity during AVHs in schizophrenia patients, in order to check whether inner speech occurred.

*Methods:* Lip muscle activity was recorded during covert AVHs (without articulation) and rest. Surface electromyography (EMG) was used on eleven schizophrenia patients.

*Results:* Our results show an increase in EMG activity in the orbicularis oris inferior muscle, during covert AVHs relative to rest. This increase is not due to general muscular tension since there was no increase of muscular activity in the forearm muscle.

*Conclusion:* This evidence that AVHs might be self-generated inner speech is discussed in the framework of a predictive control model. Further work is needed to better describe how the inner speech monitoring dysfunction occurs and how inner speech is controlled and monitored. This will help better understanding how AVHs occur.

*Introduction*

Auditory verbal hallucinations (AVHs) are a common feature of the schizophrenia psychosis, affecting 50% to 80% of the patients (Andreasen, & Flaum, 1991; Nayani and David, 1996). Hallucinations are defined as "a sensory experience which occurs in the absence of corresponding external stimulation of the relevant sensory organ, has a sufficient sense of reality to resemble a veridical perception over which the subject does not feel s/he has direct and voluntary control, and which occurs in the awake state" (David, 2004, p. 110). AVHs or "hearing voices" can thus be considered as speech perception in the absence of a relevant external acoustic input. AVHs are often distressing and engender suffering, functional disability as well as social marginalization (Franck, 2006). They are a complex and non-unitary phenomenon that can be explained by different underlying mechanisms (Larøi & Woodward, 2007).

Many theoretical models have been proposed to explain AVHs in schizophrenia (see e.g. David, 2004, for a review). A widely cited and influential model formulates AVHs as a dysfunction of the monitoring of inner speech (Feinberg, 1978; Frith, 1992). Inner or covert speech can be defined as internal non-vocalized, non-articulated speech, produced in one's mind (MacKay, 1992). It has been shown to activate a similar network of cerebral regions as overt speech (McGuire, Silbersweig, Murray, David, Frackowiak, & Frith, 1996; Shergill, Bullmore, Brammer, Williams, Murray, & McGuire, 2001). The model claims that, due to a failure of the self-monitoring mechanism, the inner speech of the patient is not identified as self-generated and is experienced as coming from an external source, i.e., as external voices. In other words, the model posits that the mechanism for distinguishing self-generated from externally-generated percepts is deficient.

Specifically, it has been suggested that inner speech may share some properties with motor actions (Feinberg, 1978; Frith, 1992; Jones & Fernyhough, 2007). Although the hypothesis that "inner speech is a kind of action" (Gallagher, 2004, p. 11; Jones & Fernyhough, 2007, p. 396) is still a matter of debates (see e.g. Vosgerau & Newen, 2007) it has led to interesting theoretical modelling, via the motor control framework (Frith, 1996) and typically via the 'predictive model' (also called 'comparator model'). In this framework, it is suggested that the brain uses forward modelling to predict the consequences of action (Wolpert, Ghahramani, & Jordan 1995; Wolpert, 1997). When motor commands are sent to the motor system to achieve a particular intended end-state, an efference copy is issued in parallel. This efference copy is used to calculate a prediction of the sensory state of the motor plan.

As illustrated in Figure 1, the predictive model includes two internal models (Kawato, Furukawa & Suzuki, 1987; Wolpert, 1997; Wolpert, et al., 1995): (a) an *inverse model* which computes motor commands based on the intended action and (b) a *forward model* which predicts the sensory consequences from a copy of the motor commands. The inverse model transforms the intention of action into motor commands, which are sent to the motor system and generate movement. A copy of the motor commands is simultaneously sent to the forward model, which generates a prediction of the sensorimotor consequences of the motor commands.

---
PLACE FIGURE 1 AROUND HERE
---

The predictive model has many advantages. First, with a predictive system, errors can be detected before sensory feedback reaches the central nervous system. The propagation of the sensory feedback to the central nervous system is not instantaneous, due to axon transmission times and synaptic delays. The predictive system can overcome this problem. The forward model predicts the consequences of implementing motor commands in terms of kinematics and sensations. If we can predict where and when our tongue is going to be after the motor command has been issued, then we can check whether the correct command has been selected before we actually make the movement. This internal feedback is available much more rapidly than the actual feedback signals resulting from the movement. Motor commands can therefore be adjusted online to reach the intended goal (integrating the error in the comparison between predicted state and intended state, see Figure1), without having to rely on delayed sensory feedback, which makes for smooth motor actions (Kelso, 1982; Miall & Wolpert, 1996).

Secondly, the predictive system provides information about the source of sensations, or agency. It is suggested that if the actual sensory feedback matches the predicted state then self-authorship is experienced. Frith (1996) claims that a defective predictive system could explain why a self-initiated action may be experienced as controlled by someone else in the 'delusion of control' symptom: if the predicted sensory feedback (corollary discharge) does not match the actual sensory feedback then some external influence must have been at work.

A further advantage of the predictive model is that when actual and predicted states match, the sensory effect of the motor act is modulated, attenuating it perceptually compared with identical stimulation that is externally produced (Blakemore, Frith, & Wolpert, 1999; Blakemore, 2003; Frith, 2002). This suggests that a modulation/attenuation of sensory cortex activity occurs when an action is self-initiated. In other words, if we can predict the sensations we are going to feel then they are not needed and they can be attenuated.

Forward and inverse models can be improved or adapted on the basis of the errors resulting from the comparison between the actual sensory experience and the predicted and intended states respectively (see Figure 1). Although the use of internal models of sensory expectations in motor control is a fruitful hypothesis in motor control research and specifically in speech monitoring (Guenther, Ghosh & Tourville, 2006; Houde & Nagarajan, 2011; see also Postma, 2000) some researchers have provided data that question the fact that internal models can be adapted or generalised. Tremblay, Houle & Ostry (2008) for instance, show no generalisation of adaptation to altered jaw dynamics during word production, contrary to the predictions of control based on internal model improvement or generalisation.

Predictive control modelling of movement (including speech) has been suggested to also apply to inner speech or conscious thought (Feinberg, 1978, Frith, 1996). Implication of a corollary discharge mechanism in inner speech control is supported by several neuroimaging studies demonstrating a frontal modulation of the temporal regions during inner speech production (Ford & Mathalon, 2004; Paus, Perry, Zatorre, Worsley & Evans, 1996 (for silent speech, without vocalization but with articulation); Shergill et al., 2002). In addition, the involvement of a motor control system during inner speech production is also supported by empirical research suggesting that motor commands are emitted. Subtle muscle activity has been detected in the speech musculature using electromyography (EMG) during verbal mental imagery, silent reading and silent recitation (Jacobson, 1931; Livesay, Liebke, Samaras, & Stanley, 1996; McGuigan & Dollins, 1989). This suggests that motor commands are generated during inner speech production in healthy participants.

AVHs in schizophrenia can be explained in the framework of predictive control of inner speech, as proposed by Seal, Aleman & McGuire (2004) as well as Jones & Fernyhough (2007) (see Feinberg, 1978 for an earlier similar account). These authors suggest that, during AVH, because of a distortion or absence of a predicted state, the actual sensory consequences of the inner speech are not attenuated. Either because of attributional biases (Seal et al., 2004), or because the emotion of self-authorship is not felt (Jones & Fernyhough, 2007), inner speech is then experienced as other-generated. Therefore, a dysfunction in the predictive model would result in the experience of an AVH.

Neuroimaging studies of schizophrenia patients have provided evidence for the suggestion that the corollary discharge from the frontal speech areas fails to inform the temporal lobes that the auditory input is self-generated. Disrupted fronto-temporal pattern of cerebral activation has been shown in schizophrenia patients relative to healthy participants (Ford & Mathalon, 2005; Lawrie et al., 2002; McGuire et al., 1995; Rapin et al., 2012).

Within this framework, i.e., if AVHs are the results of inner speech not being identified as self-produced due to a predictive control deficiency, then AVHs should be associated with inner speech production. Therefore, the presence of speech-related motor commands during AVHs would confirm that AVHs are self-generated inner speech production. If they exist, these motor commands could result in very small and visually undetectable activity of the orofacial speech muscles. Several studies have tried to examine orofacial muscular activity during AVHs.

Gould (1948, 1949) measured the myoelectric activity of the vocal musculature (lower lip and chin) during a resting period in patients with and without AVHs as well as in healthy participants. He concluded that AVHs are associated with activity of the speech musculature. Nevertheless, he did not control for the presence *vs.* absence of AVHs and he also observed high amplitude muscular activity in 28% of the control participants, suggesting that the recorded activity might in fact have corresponded to a generalized muscular tension and not to AVHs. In one of the patients, subvocal speech (slight whisper) was also present during AVHs, and the content of the audio recording corresponded to the patient's report of the voices' content. Therefore, Gould's findings are in line with the account of AVH as self-generated soft (or quiet) speech, but not exactly inner speech. The activity measured by Gould could in fact correspond to a subtype of AVHs, produced with subvocalization. AVHs accompanied by movements of the lips and tongue have been described as a specific form of AVHs (Séglas, 1892) but many hallucinating patients do not subvocalize or articulate. Therefore Gould's finding could be specific to a particular type of hallucinations. Few studies tried to replicate Gould's observations. McGuigan (1966) studied one patient who signalled his AVHs by pressing a button at the start of a voice until the end of its occurrence. McGuigan observed a significant increase in the chin muscles activity two seconds prior to an AVH report but not six seconds before. Crucially, the increase in the chin muscle activity was accompanied by a faint whisper. Inouye and Shimizu (1970) also reported increased invasive EMG activity from the speech musculature in nine hallucinating patients in 47.6% of their hallucinations. Similar to McGuigan's findings, they observed a muscular activity increase prior to the start of an AVH. Nonetheless, the authors did not specify whether AVHs were accompanied by subvocalization. In the same theoretical framework, Bick and Kinsbourne (1987) found that patients report less AVHs when they are asked to perform a motor task involving the mouth (vs. control motor tasks).

In parallel with these few findings in which AVHs seemed to be associated with EMG activity, contrary results were also found. Junginger and Rauscher (1987) recorded vocal and non-vocal invasive EMG activity in 19 hallucinating and 22 non-hallucinating psychiatric patients. The authors failed to find neither a relationship between increases in vocal EMG and reports of hallucinations nor an acoustic subvocal activity. Green and Kinsbourne (1990) did not find EMG activity during AVHs either. These results suggest that when AVHs are not subvocalized they would not systematically involve speech muscle activity. This suggestion is supported by a review by Stephane and colleagues (2001).

In sum, some studies have suggested the presence of orofacial muscular activity during AVHs. The findings were not systematically replicated, however, making it impossible to conclude on a speech-related muscular activity during AVHs. These studies had important methodological limitations (limited number of participants, lack of clear correlation between EMG measurements and AVH reports, imprecision in the time-windows taken into account for analysis). Moreover, most of the studies concluding positively on a muscular activity associated with AVHs either also clearly showed the existence of a subvocal activity or did not specify whether the muscular activity was accompanied by subvocalization. This suggests that these positive results could be valid only for a specific type of hallucinations. Therefore it seems crucial to reproduce muscular recordings in schizophrenia patients suffering from non-subvocalized AVHs in order to check that the inner-speech-monitoring-dysfunction account is suited for all types of AVHs.

The question whether "inner" AVHs result in myoelectrical activity still holds, as the EMG studies of AVH provide conflicting results. The aim of this study was thus to examine lip muscle activity during the occurrence of AVHs in schizophrenia patients, without co-occurrent subvocalization, using a replicable protocol. If the inner-speech-monitoring-dysfunction account is correct, then we predicted that lip muscle activity should be higher during AVHs than during a resting state without AVHs.

Because of the exploratory nature of this study, no average measurements were known beforehand, making it impossible to predict magnitude effects and to design for a power analysis. An emphasis was put on designing an efficient and suitable protocol for the patients. We were interested in collecting data from patients to obtain objective measurements on speech muscular activation during AVHs as well as to derive an order of magnitude usable in further studies.

## *Methods*

**Participants**

14 schizophrenia patients (mean age=37.8, standard deviation (SD) =12.1) participated in this study. Three participants were discarded because the symptom of AVH was not confirmed. The group of 11 patients analyzed (mean age=37.17, SD=12.5) included five women. Eight patients were right-handed, two were left-handed and one was ambidextrous (Oldfield, 1971). All patients were native speakers of French. Under the DSM-IV-TR criteria (APA, 2000), 10 patients were diagnosed paranoid schizophrenic and one undifferentiated schizophrenic. All patients were receiving antipsychotic medication and were experiencing frequent hallucinations at the time of the study (as assessed by an AVH questionnaire). Patients presenting language and/or auditory deficits were excluded from the study. Patients were recruited at the *Centre Hospitalier Universitaire* of Grenoble (Grenoble university hospital) and gave written consent to participate in the study, which was approved by the university and hospital ethics committee of Clinical Research (CPP-09 CHUG-17).

**Preliminary Study**

To measure possible muscular activity associated with the beginning of an AVH, it was necessary to know when the phenomenon happened by asking the patient to signal its occurrence. Obviously, the patient cannot report her/his AVHs orally since this would result in an artefactual activation of the speech musculature. A number of previous studies (McGuigan, 1966; Inouye & Shimizu, 1970) used a button pressing method to signal AVHs. The patients pressed a button once they heard a voice and kept it pressed until the voice ended. This method was used in our experiment. In order to take into account the possible delay between hearing an AVH and the report of this perceptual experience, it was necessary to estimate the button pressing reaction time to hearing speech. In order to do so, six schizophrenia patients (mean age=36, SD=13.9) from our sample additionally underwent a behavioral experiment examining this reaction time. A speech sequence consisting of a combination of six isolated words and six full sentences, separated by silences of random durations, was played through loudspeakers. Patients were instructed to press the button when they heard an utterance and to keep it pressed until the end of the utterance. Results showed a reaction time to hearing speech of 501 ms (SD=385ms). This finding was taken into account in the data processing to label the beginning of a hallucinatory period (see section on data processing).

**Setup for recordings**

Data were recorded with a Biopac MP150 surface EMG system, using the Acqknowledge software (www.cerom.fr). Bipolar surface electromyography (sEMG) recordings were obtained from two orofacial speech muscles: orbicularis oris superior (OOS) and orbicularis oris inferior (OOI). The orbicularis muscles were chosen because they showed the most activation in the EMG studies presented in the introduction. Figure 2 shows the electrodes placed on the two lip muscles. A muscle in the forearm (brachioradialis) served as a control to ensure that the measured orofacial muscular activations were associated with speech and not with a generalized muscle contraction. The forehead muscles were not used because of their role in emotion-related face motions, which are common in patients experiencing AVHs.

The experiment was video-monitored using a Canon video camera to track visible facial movements. A microphone was placed 15cm away from the patient's lips to record the patient's speech during the overt reading conditions and to monitor any faint speech production during AVHs. sEMG and audio data were all recorded through the Biopac acquisition system and were thus synchronized. The recorded signals were: the muscular activities (sEMG electrodes), the audio signal, the button pressing signal and the e-prime stimuli presentation signal when used (see hereafter). These signals were recorded at the following sample frequencies: muscles: 3125Hz; audio: 25000Hz; button: 3125Hz; e-prime: 3125Hz. The experiment took place at the Grenoble university hospital.

---

PLACE FIGURE 2 AROUND HERE

---

**Procedure**

A training session was first executed and then two conditions (control and test) were run.

In the *training* condition, patients were asked to read aloud isolated words and sentences at a comfortable rate. This condition lasted approximately five minutes and was used to familiarize the patient with the setup, and to check the electrodes were well positioned on the orbicularis oris muscles.

In the *resting condition*, patients were asked to remain silent and not to move for 10 seconds. This condition was used to obtain a baseline measurement of muscle activity.

In the *hallucinatory condition*, patients were asked to remain silent, not to move, and not to prevent voices from arising. They were instructed to signal when they heard a voice by pressing a button[a] until the end of the utterance. They were told not to verbally interact with the voice, in order to limit the occurrence of subvocal speech associated with their own response to the voice. This condition lasted approximately 15 minutes, depending on the patient's responsiveness and degree of anxiety.

Our expectation was to find higher sEMG activations in the hallucinating state than in the resting state (in which quasi-constant baseline activation was expected).

The patients underwent a pre-experiment questionnaire on the frequency, latency, nature and content of their everyday AVHs and a post-experiment questionnaire inspired from the Hoffman AVH rating scale (Hoffman et al., 2003), on AVHs experienced during the hallucinatory condition.

The experiment thus unfolded as follows: pre-experiment questionnaire, silent condition, training condition, silent condition, hallucinatory condition, silent condition, post-experiment questionnaire.

---

[a] In a first version, patients were asked to press a button as soon as they heard a voice and to press it again at the end of the hallucinatory fragment (sentence, word or discourse).

**Data Processing**

With regards to the hallucinatory periods, audio, video and EMG signals were used to eliminate any AVHs with subvocalizations. More specifically, AVHs in which sound or visible muscular activity was detected (in the audio, video or EMG signals) were excluded from the analysis. If either the audio recording showed any lip or mouth noise, or if the video signal showed any lip or mouth movement, or if the EMG signal contained a visually obvious burst of activity corresponding to lip contraction, that part of the signal was excluded. This represented less than 5% of the data. Therefore, the AVHs data analyzed consisted of non-subvocalized AVHs occurrences only.

Data were exported from the Acqknowledge software to a Matlab format (www.mathworks.fr. 9.0). EMG data were pre-processed as follows. A 50Hz frequency comb filter was applied to eliminate power noise. A [10Hz-300Hz] bandpass filter was then applied in order to focus on the orofacial sEMG frequency band (DeLuca, 2002). The filtered EMG signal was then centered by subtracting its mean.

The analysis of the muscle activity was performed on specific time windows focused on the tasks. In the hallucination condition, a window starting 500ms before each button press was chosen to take the reaction time into account (see above). We had originally planned to set the end of the analysis window at the time of button release, *i.e.* at the end of the hallucination as signaled by the patient. However after examining several patients, we were not sure of the ability of the patient to rapidly signal the end of a hallucination. If, say, the hallucination began with several sentences, then the patient would wait after the last hallucinatory sentence, to make sure another one was not coming. This would make the hallucinatory period longer than it really was and possibly lead to the presence of potential interferences such as swallowing (visible muscle activity) or self-intended verbal thinking (non-visible muscle activity). Moreover, for some patients, it appeared that AVHs were so frequent that they maintained the button pressed at all times even if there were some short breaks between AVHs. We therefore restricted the analysis to the beginning of the AVHs to make sure the analysis window only contained hallucinatory events. A time window of [-500, 1500] ms around the button press thus corresponded to an interval where an AVH had most certainly occurred. Some patients nevertheless experienced AVHs lasting less than 2,000 ms and when this was the case, the actual time of button release was used for the end of the time window. This analysis strategy leads to parts of the AVH data not being used since the long AVHs were not fully taken into account (only their beginnings).

For the baseline, it was finally impossible to use the signals from the resting condition since, upon asking the patients whether they had experienced AVHs during these periods, most of them answered positively. In order to have baseline signals without AVHs, the rest signals were manually detected using parts of the signals recorded during the hallucinatory condition in which there was no button pressing *i.e.* no reported AVHs. These signals will be referred to as "Rest condition", even though in most cases they were not recorded during the rest period. These Rest signals were chunked into two-second periods to match the AVH time window length. We took the same number of Rest and AVH trials. The Rest trials were picked in a time window close to an AVH in order to maintain physiological state homogeneity.

Once each trial was isolated for each condition and each participant, the maximum of the absolute value of each pre-processed sEMG signal (one per muscle) was computed in the corresponding time window using a 200ms sliding window. Maximum activation was chosen because any lip muscle activity observable in the AVHs should be of extremely low amplitude (since no movement of the lips were involved), and because lip muscle activity is not constant, even in overt speech. Therefore the potential bursts of activation could be masked by noise, if an average activation pattern were calculated over each period. The maximum of the absolute value extracted was then transformed with a natural logarithm. The level of the EMG signal measured can be considered to result not only from muscular activity but also from skin conductivity and other factors that differ for each participant. If these factors (supposedly constant for each participant) are considered as multiplicative, the logarithmic transformation makes it possible to overcome them in the statistical analysis by making the model additive and therefore insensitive to this constant. It also served to smooth the edges of the distribution and reduced the influence of potential outliers on the statistical results.

A repeated-measures analysis of variance (ANOVA) was then applied to the mean activation measures over all patients with *Condition* (two levels: AVH, baseline) and *Muscle* (three levels: OOS, OOI, BR) as within-subjects factors. The analysis was carried out with SPSS 18.0 software (http://www.spss.com/fr/).

### *Results*

### *AVH reports*

All participants experienced covert AVHs during the experiment. Overall, the number of "voices" heard was one for just over a third of the patients and more than five for just under a third of the patients. The voice was heard clearly equivalent to an external voice by 27% of the patients, and was reported as producing an intense sense of reality to them. The voices uttered short sentences in more than half of the cases. The content of some was reported as follows: "*Arrête, arrête maintenant*" ('Stop, stop it now'); "*Tu es stupide*" ('you are stupid'); "*Aide-moi*" ('help me'); "*Arrête maintenant, ça va agir sur ton cerveau*" ('stop it now, this is going to influence your brain'); "*accident de voiture*" ('car accident'); "*âme, âme, pars au large*" ('soul, soul, go offshore'). The voice's influence on the patient was moderate, probably due to the experimental setup the patients were put in. They reported they were able to abstract themselves from the voice's content.

### *EMG results*

First, to check that the electrodes were well positioned and that only the electrodes on the lips were associated with speech, we compared EMG signals measured during the training condition with those measured during the experimental conditions. As expected, we found that lip muscle EMG activations were significantly higher in the (overt speech) training condition than in the other two conditions. In addition, we found that arm muscle activation was not higher in the training condition than in the other two conditions. After this preliminary assessment, the training condition data were not included in the analysis.

Table 1 provides the mean values and standard deviations of the peak activation for each muscle in each of the conditions of interest (AVH and Rest) for each participant. Note that the muscle activations in the Rest condition were not null, due in part to physiological noise (Tassinary, Cacioppo & Vanman, 2000). Figure 3 provides the mean of the peak activation in the AVH and Rest conditions over all patients for the orbicularis oris superior, the orbicularis oris inferior, and the brachioradialis. It can be seen that orbicularis oris inferior activation is stronger in the AVH condition than in the Rest condition, whereas the brachioradialis activation remains constant across conditions.

---
PLACE TABLE 1 AROUND HERE
---

The three by two repeated-measures ANOVA revealed a significant effect of *Condition* ($F(1,10)=11.54$, *p*<.007) as well as a significant interaction between *Condition* and *Muscle* factors ($F(2,20)=4.24$, $p<.029$). The *Muscle* effect was non-significant ($F(2,20)=3.11$, $p=.067$). In average, across all patients, the activation values were higher for the three muscles in the AVH condition than in the Rest condition (see table 1). However, the statistical interaction between *Condition* and *Muscle* suggests that this difference is not significant for all muscles. Univariate Student t-tests were therefore performed to examine the effect of condition on each muscle separately.

### **Orbicularis Oris Superior**

The Student t-test showed no significant difference of activation between AVH and Rest for the OOS ($t(10)=-1.65$, $p=.13$). In other words, the activation values of the OOS were not statistically significantly higher in the AVH condition than in the rest condition (figure 3, table 1).

*Orbicularis Oris Inferior*

The Student t-test showed significant difference of activation between AVH and Rest for the OOI ($t(10)=-3.48$, $p=.006$). This suggests an OOI activation increase in the AVH condition compared to the Rest condition (figure 3, table 1).

*Brachioradialis*

The Student t-test revealed no effect of condition ($t(10)=-1.54$; $p=.155$), as shown in figure 3 and table 1.

-------------------------------------------------------------------------------------------------------
PLACE FIGURE 3 AROUND HERE
-------------------------------------------------------------------------------------------------------

*Discussion*

This study examined the potential activation of the speech musculature during non-subvocalized AVHs in 11 schizophrenia patients using sEMG under two conditions: during auditory verbal hallucination (AVH) and during rest. Two lip muscles involved in speech production were examined —orbicularis oris superior (OOS) and inferior (OOI)— as well as one control muscle in the forearm (BR). A condition effect was revealed suggesting that, for the three muscles, the AVH condition resulted in higher muscular average activation values than the rest condition. However the analysis of the interaction between the muscles and the condition revealed that the activation increase from rest to AVH is only significant for one speech muscle, namely the OOI. Thus, AVH is associated with a lower lip muscle activity that is significantly higher than at rest. This suggests that AVHs are associated with speech specific labial motor commands. These labial motor commands could be associated with self-generated inner speech. The fact that the activity of the brachioradialis did not significantly vary across conditions confirmed that the increase in muscular activity measured in the OOI speech muscle is actually related to speech and not to general muscular tension. The difference between the two orbicularis oris activation patterns can be explained by muscle anatomy: the sEMG electrode positioned on the OOI site could in fact capture activation of muscular fibers from other muscles involved in speech, entangled with those of the OOI, such as the depressor angular muscle (Blair & Smith, 1986) or the mentalis (Nazari, Perrier, Chabanas, & Payan, 2011). This would increase the global level of activation in the lower lip, attributed to the OOI, as compared to the OOS during speech. Thus, by its proximity with other speech muscles, the OOI site better reflects muscular activity during speech production. This finding is consistent with the literature describing OOI activity (rather than other orofacial muscle) during speech and inner speech tasks (Livesay et al., 1996).

The high inter-subject variability observed in this study can be explained by the difficulty in measuring precise muscular activity using surface electrodes due to intrinsic morphological differences. Skin thickness, subcutaneous fat and facial hairs were among the non-controllable factors that influenced conductivity. Moreover, electrode placement, ambient temperature and humidity were also factors producing variability. Another factor which could potentially increase inter-subject variability is the diversity of the participants. The patients' age ranged from 19 to 61 years and the illness duration ranged from 3 to 26 years. All patients received antipsychotic medication but the treatment varied between participants. Early-onset schizophrenia is associated with a significant clinical severity and poor response to antipsychotic treatment. Age may have an impact on the patient phenotype and may affect clinical response and susceptibility to side effects (Kumra et al, 2008). Moreover, the antipsychotic/anti-hallucinatory effect, based on the antidopaminergic effect of the medication, varies depending on the affinity for dopaminergic D2 receptor and the dose of antipsychotics (Kapur & Remington, 2001). The post-experiment questionnaire did reveal some differences between participants in the properties of their hallucinations. Over a third of the patients only heard one voice during the hallucinatory periods, but some heard several voices, more than 5 voices for under a third of the patients. About a third of the participants judged the voices they heard as very clear and real. And for more than a half of the participants, the AVH consisted of very short sentences. These differences could be related to differences in age, illness duration or medication. No correlation could be found, however, between the degree of EMG activity in each patient and any of the hallucinatory properties reported by each patient in the post-experiment questionnaire, nor any of the medication. Future works should address the issue of participant diversity. Nevertheless, the statistical significance of the results obtained in this study, despite the participant diversity, is evidence for their robustness.

Our findings suggest that the hallucinatory crisis is associated with a surface electromyographic activity increase in the OOI muscle compared to baseline (i.e. rest with no AVHs). This indicates that there would be subtle speech-related muscular activity during covert AVHs providing evidence that covert AVHs could involve self-generated inner speech. This result is of importance as it provides support to the inner-speech-monitoring-dysfunction account of AVHs (Frith, 1992; Jones & Fernyhough, 2007; Seal, Aleman & McGuire, 2004). In the framework of a motor control model based on predictive control, the lower lip activity observed during AVHs could well reflect the emission of motor commands associated with the generation of inner speech. As hypothesized by the inner-speech-monitoring-dysfunction account of AVHs, in schizophrenia patients, during inner speech, motor commands would correctly be sent to the speech muscles, but the verbal self-monitoring mechanism, or the efference system, would dysfunction, resulting in inner speech being misattributed to an external agent. Our results show that AVHs are indeed associated with increased speech-related muscular activation and support the assumption that a deficit in the monitoring of inner speech is at the origin of AVHs.

As noted by Vanderwolf (1998), the neurophysiological source of this low amplitude EMG activity (during imagined action) has not been elucidated. It has been suggested that part of it arises from intrafusal muscle fibers (Malmo, 1975). Jeannerod & Decety (1995) have also suggested a mechanism by which residual EMG may be present during action imagery. According to them, during mental simulation of an action, "it is likely that the excitatory motor output generated for executing the action is counterbalanced by another parallel inhibitory output. The competition between two opposite ouputs would account for the partial block of the motoneurons, as shown by residual EMG recordings and increased reflex excitability" (p. 728). In the same vein, Rossini et al. (1999) showed that mental simulation of finger movement increases excitability of the spinal motoneuronal pools governing muscles involved in finger movement. Furthermore, a recent examination of hand-related motoneuron activity has shown that a mental arithmetic task affects the rate and variability of the tonic discharge of motor units (Bensoussan et al., 2012). The increases observed in hand EMG activity were consistent with the modulation of the motor unit discharge rate induced by mental arithmetics. This finding suggests that mental arithmetic may influence the state of the motor system, including in its most peripheral spinal component, i.e. the motoneuron. More research is necessary to better understand how (and at which neurophysiological level, cerebral, spinal, peripheral) the motor commands associated with inner speech are inhibited but may still result in residual EMG activity.

To summarise, our findings suggest that EMG activity is present in speech muscle during inner speech and are coherent with the view that "inner speech is a kind of action". Motor commands issued during inner speech, could modulate motoneuron activity related to the lips and could result in the observed subtle increase of EMG activity.

Although the inner-speech-monitoring-dysfunction account of AVHs has been extensively described and cited (Frith, 1992; Jones & Fernyhough, 2007; Seal et al., 2004), several crucial parts of it remain unexplained by its proponents. As mentioned in the introduction, the account was based on a speech motor control model involving predictive control applied to inner speech. Specifically, the account relies on the deficient comparison between the corollary discharge (or predicted sensory feedback) and the actual sensory-motor feedback leading to a lack of attenuation of the sensory consequences of self-produced actions. This would make self-produced actions, therefore, indistinguishable from externally generated sensations.

The application of the predictive control account to inner speech is problematic, however. As questioned by Frith (2012) himself, it is not clear what the actual sensory consequences of inner speech are. Inner speech generates neither kinematics nor auditory sensations, since it is internally generated. So the comparison between the predicted state and the actual state is irrelevant in the case of inner speech. We suggest two alternative interpretations.

First, our understanding of inner speech is that motor commands are generated, that are sent to appropriate speech muscles. Inhibitory signals may be sent in parallel to prevent the intensity of the motor commands from reaching a sufficient threshold for movements of the speech organs to occur. But even though the speech organs may not move during inner speech, the presence of motor commands in the speech muscles could slightly increase muscle tension and could correspond to detectable proprioceptive feedback. This suggests that in the instance of inner speech production the actual sensory feedback would in fact be a proprioceptive sensory feedback. This residual proprioceptive feedback could then be the sensory feedback that lacks a corresponding prediction during AVH and that becomes reinterpreted as an external voice.

The second possible interpretation is that during inner speech, motor commands may be sent to the motor system, but they would be inhibited and hence irrelevant as no auditory output would be produced. A copy of these motor commands would also be sent to the forward model, the output of that being the auditory voice we hear in our head (inner speech has sensory qualities and can be associated with rich suprasegmental acoustic representations, cf. Yao et al., 2012). We suggest that the relevant comparison for agency monitoring during inner speech is not that between predicted and actual sensory states but instead, the one between the intended (the input to the inverse model) and the predicted (the output of the forward model) states (the top comparator on Figure 1). Then the AVH symptoms could be explained as follows: if a deficit arises somewhere in the efference copy system, there is no match between predicted and intended states and the inner voice (predicted output) could feel alien.

The presence of EMG activity during AVH has a further possible interpretation, in agreement with mirror system or interaction theories of speech production/perception (Fadiga et al., 2002; Schwartz et al., 2010) or with the Motor Theory of speech perception (Liberman & Mattingly, 1985). The motor activity observed during inner speech could simply be an epiphenomenon of a sensory (auditory) processing of the inner voice. Any motor cortex activity or EMG activity recorded during AVH could in fact reflect the motor system resonance with the auditory images evoked by the AVH. The measurement used in our experiment cannot distinguish between this interpretation and our interpretation that AVH is a kind of action. We plan further cerebral imaging experiments to derive fine temporal analyses of cortical activity during inner speech production, which will allow us to examine whether motor cortex activity arises before sensory cortex activity (i.e. inner speech is a kind of action with sensory consequences) or the reverse (i.e. inner speech is a sensory experience with motor resonance).

Another part of the model that deserves further description is how "other" voices can be heard. Schizophrenia patients often report that they can precisely identify the voice they hear as being clearly that of someone they know and as addressing them in the second-person (Hoffman, Fernandez, Pittman, & Hampson, 2011). Similarly, inner speech can involve interior dialogues with oneself and other people, their imagined voices being clearly distinguishable from that of oneself. This suggests that inner speech can take several forms, own-inner-speech and imagined-other-inner-speech. Several studies have examined the behavioural and neural correlates of imagining speech in the voice of another person. Johnson et al. (1988), for instance, had participants listen to words spoken by a given speaker and imagine words in their own voice or in the speaker's voice. Participants were then asked to then to discriminate words they had heard from words they had imagined. Their accuracy in discriminating between perceived and produced items was higher when had imagined the words with their own voice than with the speaker's voice. This is evidence that participants can successfully produce inner speech in the voice of someone else. Mc Guire et al. (1996) examined the neural correlates of inner speech (inner generation of short sentences) and of auditory verbal imagery (imagining sentences being spoken to them in an another person's voice) using positron emission tomography. Own-inner-speech and imagined-other-inner-speech were both associated with increased activity in the left inferior frontal gyrus. In addition, imagined-other-inner-speech was associated with increase in the left premotor cortex, the supplementary motor area and the left temporal cortex, suggesting more involvement of speech perception processes, when mentally using another person's voice. Shergill et al. (2001) used fMRI to examine own-inner-speech and imagined-other-inner-speech. Imagined-other-inner-speech was associated with activation in the areas engaged during own-inner-speech task (namely left inferior frontal/insula region, the left temporo-parietal cortex, right cerebellum and the supplementary motor area) plus the left precentral and superior temporal gyri, and the right homologues of all these areas. This result was interpreted as a greater engagement of verbal self-monitoring during imagined-other-inner-speech.

Further studies should investigate whether different internal models are used for different voices. If different internal models are used for different voices then inducing an adjustment of one's internal model (using somatosensory feedback alteration, for instance) should not modify the other-voices models and thus should not modify (inner) production with someone else's voice. Future studies should also examine whether speech muscles are significantly activated as compared to resting when participants are asked to imagine sentences uttered by someone else.

These hypotheses remain highly speculative and demonstrate the need of further work to examine exactly what precise part of the speech motor control system is impaired; to understand what exactly are the sensations that lead to voice hearing during inner speech; to understand how own-inner-speech and imagined-other-inner-speech are controlled and monitored. Furthermore, factors such as attention, intentionality, externalization bias, and metacognitive beliefs are of importance in the genesis of AVHs and can assist in understanding why all inner speech production is not mistaken as coming from an external agent (Aleman & Larøi, 2008; Larøi & Woodward, 2007). They should be incorporated in any thorough account of AVHs.

This preliminary study has some limitations. First, the patients were not recorded during an inner speech condition. This condition could have been compared to the AVH condition and their potential similarity could have provided additional support to the hypothesis of inner speech involvement in AVH. A pilot study showed that it was difficult, sometimes impossible, to have the patients whisper. Some patients found the whispering task very difficult. Explaining the inner speech task could then prove even more difficult. In addition, in inner speech, there is no behavioral way of controlling the patient's production. Another limitation concerns the resting state, which was not optimally controlled. In fact, it is quite possible that intrusive inner speech (flowing unbidden thoughts) was also occurring during the rest periods and could thus result in lip muscle activity. Some authors have proposed relaxation techniques to minimise myoelectric signals during rest periods (Jacobson, 1931; Vanderwolf, 1998). This strategy could be used in further experiments.

Note that an alternative interpretation of our results could be that the collected muscular activities during the hallucinatory period correspond to the patients' replies to the voices and not to the voices themselves, which would also result in lip muscle activity. It has been shown that some patients suffering from AVHs develop a dialogue with their voices in order to tell them to stop for example (Nayani & David, 1996). In our study, the instructions were however to let the voices come and not to reply to them. All patients understood these instructions, which were moreover repeated throughout the condition.

In summary, we collected lip muscle activity during hallucinatory periods and rest in 11 schizophrenia patients. A significant increase in the activity of the lower lip muscle was observed during AVHs controlled for subvocalization compared to rest, suggesting that speech motor commands are emitted during AVHs, even without co-occurrent subvocalization. This result supports the hypothesis according to which AVHs correspond to self-generated inner speech, which would be mistakenly attributed to an external agent. Furthermore, it underlines the motor nature of AVHs, which can be analyzed in the framework of a speech motor control model including predictive control and is in agreement with theories accounting for AVHs as a misattribution of inner speech. Yet it questions the inner-speech-monitoring-dysfunction account in several of its crucial assumptions, which further studies should explore.

### *Acknowledgments*

This project was partially funded by a grant from the Région Rhône-Alpes Cluster 11 **"Handicap Vieillissement Neurosciences"** to H. Loevenbruck and M. Polosan. We are very thankful to Lionel Granjon (Gipsa-lab) for his assistance in data preprocessing and analysis. We thank our colleagues from Gipsa-lab, Coriandre Vilain, Alain Arnal and Christophe Savariaux for technical help. We also thank Stéphane Rousset (LPNC, Grenoble) and Jean-Luc Schwartz (Gipsa-lab) for helpful advice.

*Figures and Tables Legend*

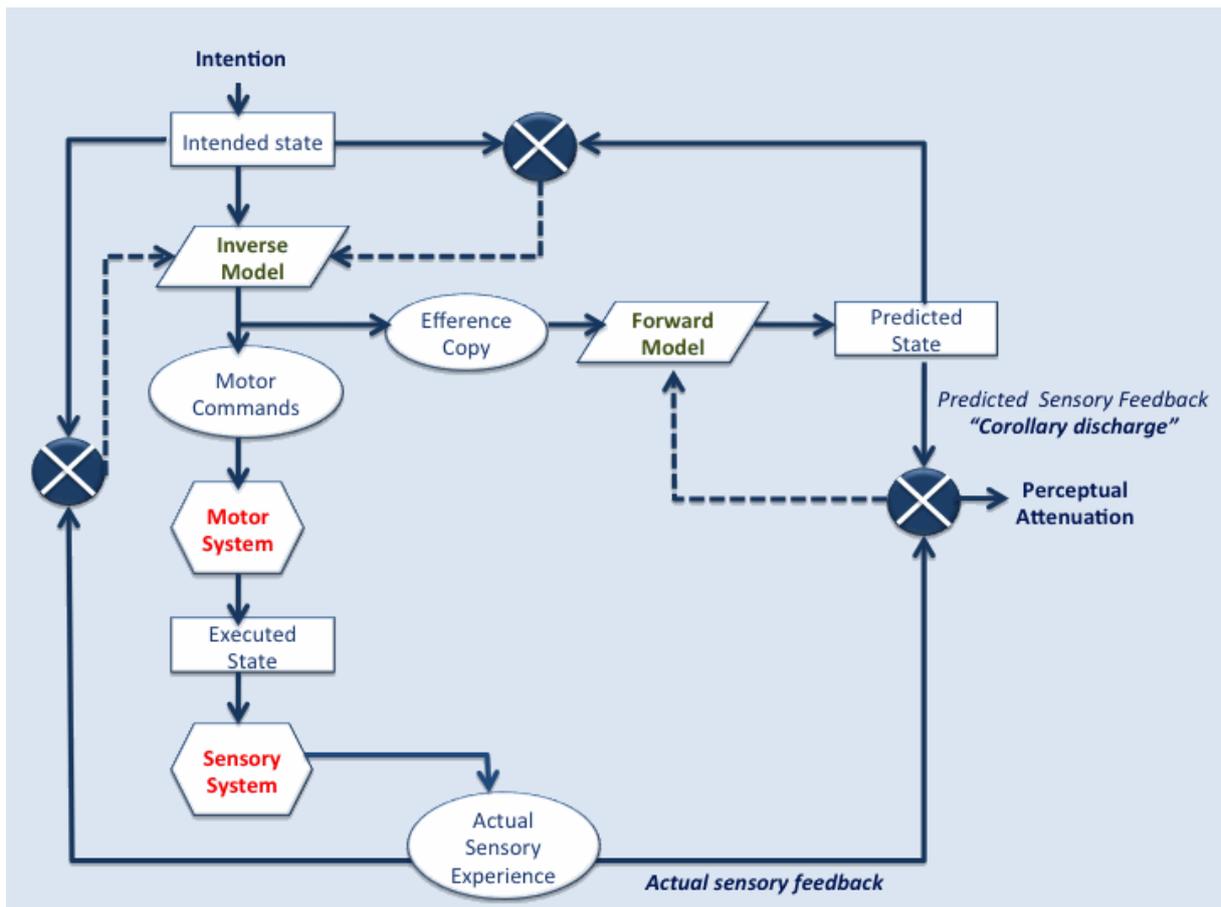

Figure 1: The Predictive Model: a forward model predicts the sensory consequences of action (estimated sensory feedback). This feedback can be used to cancel out self-generated sensory events, thus distinguishing them from sensory events with an external cause. Adapted from **"**Explaining the Symptoms of Schizophrenia: Abnormalities in the Awareness of Action,**"** by C. Frith, S. Blakemore, and D. Wolpert, 2000, Brain Research Reviews, 31, p. 359. Copyright 2000 by Elsevier. Adapted with permission.

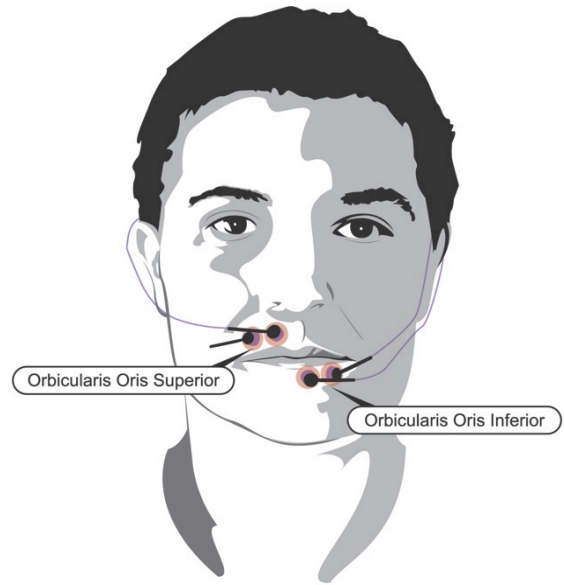

Figure 2: Surface electromyography electrodes placed on the lip muscles examined.

|  |  | P1 | P2 | P3 | P4 | P5 | P6 | P7 | P10 | P11 | P13 | P14 | Mean Value | Standard Deviation |
|---|---|---|---|---|---|---|---|---|---|---|---|---|---|---|
| Orbicularis Oris Superior | AVH | 2.36 | 4.10 | 5.38 | 3.45 | 3.36 | 2.57 | 3.60 | 4.39 | 3.35 | 2.75 | 1.74 | 3.37 | 1.020 |
| | Rest | 2.44 | 3.95 | 5.13 | 3.48 | 3.46 | 2.54 | 2.66 | 3.95 | 3.34 | 2.49 | 1.88 | 3.21 | 0.926 |
| Orbicularis Oris Inferior | AVH | 2.60 | 4.74 | 4.83 | 4.47 | 3.26 | 2.66 | 3.03 | 3.80 | 3.63 | 4.11 | 1.71 | 3.53 | 0.985 |
| | Rest | 2.27 | 3.87 | 3.73 | 4.06 | 3.33 | 2.51 | 2.10 | 3.62 | 3.68 | 3.17 | 1.55 | 3.08 | 0.839 |
| Brachio-radialis | AVH | 3.62 | 2.81 | 3.67 | 2.87 | 2.77 | 2.70 | 2.32 | 2.51 | 2.61 | 2.69 | 2.46 | 2.82 | 0.439 |
| | Rest | 3.66 | 3.01 | 3.09 | 2.60 | 2.69 | 2.60 | 2.14 | 2.37 | 2.49 | 2.99 | 2.15 | 2.71 | 0.452 |

Table 1: Mean values and standard deviations of peak muscular activation for each muscle and in each condition, for each patient.

*Note*. Patients P8, P9, and P12 were excluded because the symptom of AVH was not confirmed. AVH = auditory verbal hallucination.

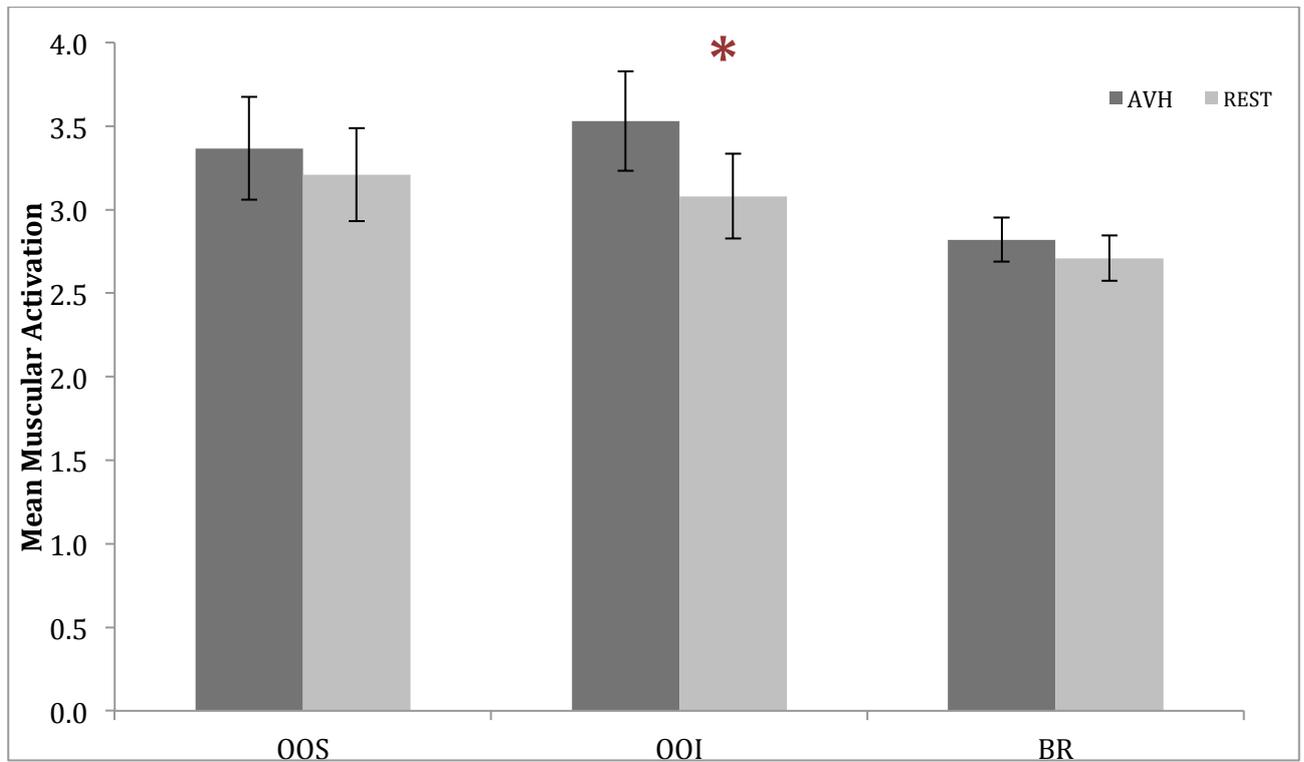

Figure 3: Means of the maxima of activation of the orbicularis oris superior (OOS), the orbicularis oris inferior (OOI) and the brachioradialis (BR) in the AVH condition (dark grey) and in the Rest condition (light grey). The error bars represent one standard error.